\documentclass[aps,pra,amsmath,amssymb,twocolumn]{revtex4}
\usepackage[T1]{fontenc}
\usepackage{amssymb,amsmath}
\usepackage{amsfonts}
\usepackage{graphicx}
\usepackage{color}
\usepackage{enumerate}
\usepackage{mathbbol}
\usepackage{mathrsfs}
\usepackage{listings}

\def\nocorset{\mathcal{SNC}}
\def\cals{\mathcal{S}}
\newcommand{\bn}{{\mathbf n}}
\def\ket#1{|#1\rangle}
\def\ketbra#1{|#1\rangle\langle#1|}

\def\bbr{\mathbb{R}}

\def\bra#1{\langle#1|}
\def\tr{\mathrm{tr}}
\def\fn{\mathfrak{N}}

\renewcommand{\leq}{\leqslant}
\renewcommand{\geq}{\geqslant}

\begin{document}

\title{Genuinely entangled symmetric states with no $N$-partite correlations}

\author{S.~Designolle$^{1,2}$, O.~Giraud$^{2}$, and J.~Martin$^{3}$}
\affiliation{
$^{1}$\'Ecole polytechnique, 91128 Palaiseau Cedex, France\\
$^{2}$LPTMS, CNRS, Univ.~Paris-Sud, Universit\'e
    Paris-Saclay, 91405 Orsay, France\\
$^{3}$Institut de Physique Nucl\'{e}aire, Atomique et de Spectroscopie, CESAM,
Universit\'{e} de Li\`{e}ge, B\^{a}t.~B15, B - 4000 Li\`{e}ge, Belgium.
}

\date{July 13, 2017}

\begin{abstract}
We investigate genuinely entangled $N$-qubit states with no $N$-partite correlations in the case of symmetric states. Using a tensor representation for mixed symmetric states, we obtain a simple characterization of the absence of $N$-partite correlations. We show that symmetric states with no $N$-partite correlations cannot exist for an even number of qubits. We fully identify the set of genuinely entangled symmetric states with no $N$-partite correlations in the case of three qubits, and in the case of rank-2 states.  We present a general procedure to construct families for an arbitrary odd number of qubits.
\end{abstract}

\maketitle

%%%%%%%%%%%%%%%%%%%%%%%%%%%%%%%%%%%%%%%%%%%%%%%%%%%%%%%%%%%%%%%%%%%%%%%%%
%%%%%%%%%%%%%%%%%%%%%%%%%%%%%%%%%%%%%%%%%%%%%%%%%%%%%%%%%%%%%%%%%%%%%%%%%
\section{Introduction}
\label{sec:intro}
%%%%%%%%%%%%%%%%%%%%%%%%%%%%%%%%%%%%%%%%%%%%%%%%%%%%%%%%%%%%%%%%%%%%%%%%%
%%%%%%%%%%%%%%%%%%%%%%%%%%%%%%%%%%%%%%%%%%%%%%%%%%%%%%%%%%%%%%%%%%%%%%%%%

Entanglement is one of the most remarkable aspects of quantum physics. A pure state is not separable, or entangled, if it cannot be written as a tensor product of single-party states, and a mixed entangled state is a state that cannot be written as a mixture of pure separable states. While the case of bipartite systems is well understood, the multipartite case is much more involved, as there are many ways in which a state can be entangled. A pure multipartite state may be fully separable, that is, a tensor product of pure states of each party. In such a case, there are no correlations between subsystems. A pure multipartite state may also be biseparable, if it can be written as the tensor product of two pure states under a certain bipartition of the system. The strongest form of entanglement  in a multipartite state occurs when the state is not biseparable whatever the bipartition. This leads to the following definition: a genuinely $N$-partite entangled pure state is a state which is entangled (not biseparable) along any bipartition. Similar hierarchies exist for mixed states. In particular, a genuinely $N$-partite entangled mixed state is a state such that in any of its pure state decompositions, there exists at least one pure state which is genuinely $N$-partite entangled \cite{Wer89}. Or, as put by Bennett {\it et al.} \cite{Ben11}, by mixing pure states which do not have genuine $N$-partite entanglement one cannot obtain mixed states with genuine $N$-partite entanglement.

In quantum physics, correlations between subsystems are central in the question of entanglement. They can lead to correlations between measurement results violating Bell inequalities, thereby discarding the possibilities of local hidden variable theories. In the case of mixed states, the distinction between classical and quantum correlations is a subtle one (see e.g.~\cite{Oll01,Hen01,Ben11}). While for pure states, correlations imply entanglement and vice versa, this is not true anymore for mixed states. It is possible to construct separable mixed states which possess quantum correlations~\cite{Oll01}. Conversely, there exist genuinely entangled $N$-qubit states with vanishing $N$-partite correlation functions~\cite{Kas08}. In \cite{Wie09,Sch15}, genuinely entangled multiphoton states with vanishing $N$-partite correlation functions were created experimentally: qubits were encoded in the polarization of photons. Entanglement in these states cannot be detected by usual multipartite Bell inequalities involving only $N$-partite correlations, which led to the construction of suitable Bell inequalities able to detect $N$-partite entanglement but involving only $(N-1)$-partite correlations~\cite{Wie12}. In \cite{Sch15}, a continuous family of genuinely entangled three-qubit states without three-partite correlations was constructed. Families of examples were obtained for any odd number of qubits in \cite{Tra17}, and an analytical construction of genuinely entangled rank-4 $N$-qubit states without $N$-partite correlations for any even number $N\geq 4$ of qubits was presented.

In this paper, we investigate the case of $N$-qubit symmetric states and present a general procedure to construct families of genuinely entangled states with no $N$-partite correlations. To this end, we use the recently introduced tensor representation of spin states~\cite{Gir15} that we briefly review in Sec.~\ref{sec:tensor}. The consequences of the absence of $N$-partite correlations is expressed in terms of these tensors in Sec.~\ref{sec:symchar}. Using entanglement criteria devised in Sec.~\ref{sec:entcrit}, we fully identify the set of genuinely entangled symmetric three-qubit states with no correlations in Sec.~\ref{sec:threequbits}, and present a general construction procedure for arbitrary number of qubits in Sec.~\ref{sec:families}.

%%%%%%%%%%%%%%%%%%%%%%%%%%%%%%%%%%%%%%%%%%%%%%%%%%%%%%%%%%%%%%%%%%%%%%%%%
\section{Mixed symmetric states in the tensor representation}\label{sec:tensor}
%%%%%%%%%%%%%%%%%%%%%%%%%%%%%%%%%%%%%%%%%%%%%%%%%%%%%%%%%%%%%%%%%%%%%%%%%

Symmetric states of $N$ qubits are pure states which are invariant under permutation of the qubits, or mixtures of such pure states. Symmetric states can be expanded in the basis of Dicke states $\{\ket{D_N^{(k)}}:0\leq k \leq N\}$ defined by
\begin{equation}
\label{DickeStates}
{\ket{D_N^{(k)}}=\frac{1}{\sqrt{\binom{N}{k}}}\sum_\pi \ket{\underbrace{0 \dots 0}_{N-k} \underbrace{1\dots 1}_{k}}},
\end{equation}
where the sum runs over all permutations of the qubits. It is convenient to introduce the projector 
\begin{equation}
\label{projps}
P_{\cals}=\sum_k\ketbra{D_N^{(k)}}
\end{equation}
onto the symmetric subspace $\cals$ spanned by Dicke states \eqref{DickeStates}.

The density matrix associated with a symmetric state can be expressed in the Dicke basis as an $(N+1)\times (N+1)$ positive semidefinite matrix of unit trace. A convenient representation for symmetric states was introduced in \cite{Gir15}. In this representation, any density matrix $\rho$ can be expressed as
\begin{equation}
\label{projrho}
\rho=\frac{1}{2^N} x_{\mu_1\mu_2\ldots\mu_N} S_{\mu_1\mu_2\ldots\mu_N},
\end{equation}
with implicit summation over indices $\mu_i$ ($1\leq i \leq N$), $0\leq\mu_i\leq 3$. Here $S_{\mu_1\mu_2\ldots\mu_N}$ are $(N+1)\times (N+1)$ Hermitian matrices defined, for $0\leqslant k,k'\leqslant N$, by their entries
\begin{equation}
\label{matrixS}
\left(S_{\mu_1\mu_2\ldots\mu_N}\right)_{kk'}=\bra{D_N^{(k)}}\sigma_{\mu_1} \otimes \cdots \otimes \sigma_{\mu_N} \ket{D_N^{(k')}},
\end{equation}
where $\sigma_{1},\sigma_2,\sigma_3$ are the three Pauli matrices (with $\sigma_0$ the $2\times 2$ identity matrix). The coordinates $x_{\mu_1\mu_2\ldots\mu_N}$ are real numbers invariant under permutation of the indices and such that
\begin{equation}
\label{Tensorrep}
x_{\mu_1\mu_2\ldots\mu_N}=\tr(\rho S_{\mu_1\mu_2\ldots\mu_N}).
\end{equation}
Note that, since $\rho$ is a symmetric state, we denote by the same symbol its $2^N\times 2^N$ representing matrix in the computational basis and its $(N+1)\times (N+1)$ representing matrix in the Dicke basis. In particular we have $\rho=P_{\cals}\rho P_{\cals}$, with $P_{\cals}$ the projector \eqref{projps}. As a consequence, using  \eqref{matrixS} and \eqref{Tensorrep}, we have $\tr(\rho \sigma_{\mu_1} \otimes \cdots \otimes \sigma_{\mu_N})=\tr(\rho P_{\cals} \sigma_{\mu_1} \otimes \cdots \otimes \sigma_{\mu_N}P_{\cals})=\tr(\rho S_{\mu_1\mu_2\ldots\mu_N})=x_{\mu_1\mu_2\ldots\mu_N}$. Therefore,  these coordinates have a physical interpretation in terms of correlators, as
\begin{equation}
\label{correl}
x_{\mu_1\mu_2\ldots\mu_N}=\big\langle \sigma_{\mu_1} \otimes \cdots \otimes \sigma_{\mu_N}\big\rangle_\rho.
\end{equation}
Since $S_{0\ldots0}$ is the identity matrix, we have in particular $x_{0\ldots0}=\tr\rho=1$. Properties of the Pauli matrices imply moreover that for any $\mu_i$
\begin{equation}
\label{traceless}
\sum_{a=1}^3 x_{\mu_1\ldots\mu_{N-2}aa}=x_{\mu_1\ldots\mu_{N-2}00}.
\end{equation}
This is due to the fact that for $N=2$, we have the identity $\sum_{a=1}^3 P_{\cals}\sigma_{a} \otimes\sigma_{a}P_{\cals}=P_{\cals}\sigma_{0} \otimes\sigma_{0}P_{\cals}$. Since the $ x_{\mu_1\mu_2\ldots\mu_N}$ are invariant under permutation of indices, the position of the two indices $a$ in \eqref{traceless} does not matter.
As has been shown in~\cite{Gir15}, this representation allows easily to express the $k$-qubit reduced density matrix $\rho_k$ obtained by tracing out $N-k$ qubits just by replacing the last $N-k$ indices of $x_{\mu_1\mu_2\ldots\mu_N}$ by zero. The expansion (\ref{projrho}) for $\rho_k$ thus reads
\begin{equation}
\label{projrhok}
\rho_k=\frac{1}{2^k} x_{\mu_1\mu_2\ldots \mu_k 0\ldots 0} S_{\mu_1\mu_2\ldots\mu_k}.
\end{equation}
Note that, because of symmetry of $\rho$, the choice of qubits traced out does not matter. For instance for five qubits, tracing out qubits 2 and 4 gives  $x_{\mu_1 0 \mu_2 0 \mu_3}=x_{\mu_1\mu_2\mu_300}$, and the coordinates of the reduced density matrix would be given by $x_{\mu_1 \mu_2 \mu_3}$, whichever pair of qubits is traced out.
For single qubit states, $S_\mu=\sigma_\mu$, and the representation (\ref{projrho}) reduces to the usual Bloch representation. In the Bloch representation, a single qubit state $\rho$ can be expressed (with implicit summation over $\mu=0,\ldots,3$) as 
\begin{equation}
\label{projrho12}
\rho=\frac{1}{2} n_\mu \sigma_\mu,
\end{equation}
where $n$ is the 4-vector given by $n=(n_0,n_1,n_2,n_3)$ with $n_0=1$. The Bloch vector associated with the state is ${\bf n}=(n_1,n_2,n_3)=\tr (\rho \boldsymbol{\sigma})$, where $\boldsymbol{\sigma}=(\sigma_1,\sigma_2,\sigma_3)$. In the case of pure states, the Bloch vector ${\bf n}=(\sin\theta\cos\varphi,\sin\theta\sin\varphi,\cos\theta)$ is of unit length and we denote by $\ket{\bn}$ the corresponding qubit state. A fully separable pure symmetric state $\ket{n}\equiv\ket{\bn}^{\otimes N}$ is the tensor product of $N$ copies of a pure qubit state $\ket{\bn}$. It can be expanded in the Dicke basis as
\begin{equation}
\label{symsep}
\ket{n}=\sum_{k=0}^{N}\sqrt{\binom{N}{k}}\left[\sin\!\frac{\theta}{2}\right]^{k}\left[\cos\!\frac{\theta}{2}\,e^{-i\varphi}\right]^{N-k}\ket{D_N^{(k)}},
\end{equation}
and in the representation \eqref{projrho} it can be written as~\cite{Gir15}
\begin{equation}
\label{projrhocoh}
\ketbra{n}=\frac{1}{2^N} n_{\mu_1}n_{\mu_2}\ldots n_{\mu_N} S_{\mu_1\mu_2\ldots\mu_N}.
\end{equation}
Fully separable states are central in the context of entanglement of symmetric states. Indeed, let $\rho$ be an $N$-qubit symmetric state that is separable along some bipartition of the qubits. Then $\rho$ is a convex combination of pure symmetric states separable along the same bipartition (see e.g.~\cite{Boh16}, Section C). But any pure symmetric state separable along some bipartition is separable along any bipartition and thus fully separable \cite{Ich08}. Therefore, in the subspace of symmetric states, separable states coincide with the convex hull of the projectors $\ketbra{n}$. In other words, symmetric states are either genuinely entangled or fully separable.

%%%%%%%%%%%%%%%%%%%%%%%%%%%%%%%%%%%%%%%%%%%%%%%%%%%%%%%%%%%%%%%%%%%%%%%%%
\section{Symmetric states with no $N$-partite correlations}
\label{sec:symchar}
%%%%%%%%%%%%%%%%%%%%%%%%%%%%%%%%%%%%%%%%%%%%%%%%%%%%%%%%%%%%%%%%%%%%%%%%%

\subsection{Characterization in terms of tensor coefficients}

Symmetric states with no $N$-partite correlations are defined in \cite{Sch15} as states $\rho$ such that $\langle \sigma_{a_1}\otimes\cdots\otimes\sigma_{a_N}\rangle_\rho=0$ for any $a_i$ with $1\leq a_i\leq 3$ (in the present paper latin indices range from 1 to 3 while greek indices range from 0 to 3). Because of \eqref{correl}, this condition can be expressed in terms of coordinates $x_{\mu_1\mu_2\ldots\mu_N}$ as
\begin{equation}
\label{nocorr}
x_{a_1a_2\ldots a_N}=0\qquad\forall a_i=1,2,3.
\end{equation}
For symmetric states, because of the relation \eqref{traceless}, the absence of $N$-partite correlations has the immediate consequence that all $(N-2k)$-partite correlations, $k=0,\ldots,\lfloor N/2 \rfloor$, vanish. If $N$ is odd, it can be expressed as the hierarchy of conditions
\begin{equation}
\label{nocorrsym}
\begin{array}{rclrcl}
x_{a_1a_2a_3\ldots a_N}&=&0&\qquad\forall a_i&=&1,2,3,\quad 1\leq i\leq N\\
x_{a_1a_2\ldots a_{N-2}00}&=&0&\qquad\forall a_i&=&1,2,3,\quad 1\leq i\leq N-2\\
&\vdots& & &\vdots& \\
x_{a_1 0 0\ldots 0}&=&0&\qquad\forall a_1&=&1,2,3.
\end{array}
\end{equation}
If $N$ is even, it leads to $x_{0\ldots 0}=0$, which contradicts the fact that $x_{0\ldots 0}=\tr\rho=1$. Thus no symmetric states with no $N$-partite correlations can exist for an even number of qubits. Similarly, if $N$ is odd, no symmetric states with no $(N-1)$-partite correlations can exist. This generalises to all symmetric states the results found in~\cite{Las12} that all correlations between an odd number of subsystems vanish (and thus admit a local hidden-variable model) in an even mixture of Dicke states for any odd number of qubits.

Note that because of Eq.~(\ref{projrhok}), the hierarchy of conditions \eqref{nocorrsym} implies that all $k$-qubit reduced states obtained from $\rho$ by tracing out an even number of qubits are also states with no $N$-partite correlations. Moreover, the last condition in \eqref{nocorrsym} can be rephrased as $\langle \boldsymbol{\sigma}\rangle_{\rho_1} =0$. Such states with vanishing expectation value of the spin have been called $1$-anticoherent states, by contrast with coherent states which maximize this expectation value. They are characterized by the fact that their one-qubit reduced density matrix is always the maximally mixed state~\cite{Bag14,Gir15}. Thus, all symmetric states with no $N$-partite correlations are $1$-anticoherent.

As we mentioned, only symmetric states with an odd number $N=2M+1$ of qubits can be such that their $N$-partite correlations are zero. In what follows, we will restrict ourselves to this odd case, and we denote by $\nocorset_N$ the set of symmetric $N$-qubit states with no $N$-partite correlations.

%%%%%%%%%%%%%%%%%%%%%%%%%%%%%%%%%%%%%%%%%%%%%%%%%%%%%%%%%%%%%%%%%%%%%%%%%
\subsection{Antistates}
\label{antistates}
%%%%%%%%%%%%%%%%%%%%%%%%%%%%%%%%%%%%%%%%%%%%%%%%%%%%%%%%%%%%%%%%%%%%%%%%%

Antistates were defined in \cite{Sch15} to construct examples of genuinely entangled states with no $N$-partite correlations. In this subsection we introduce these states, and we will use them in the next subsection to characterize elements of $\nocorset_N$. 

Let $\fn=\sigma_3\sigma_1K$ be the one-qubit universal-not operator, with $K$ the complex conjugation operator. This operator is antilinear and antiunitary, and it also satisfies $\fn^2=-\mathbb{1}$.  For any pure state $\ket{\psi}$, its antistate $\ket{\bar{\psi}}$ is defined by applying the universal-not operator on each qubit, namely $\ket{\bar{\psi}}=\fn^{\otimes N}\ket{\psi}$. The antistate $\ket{\bar{\psi}}$ is orthogonal to $\ket{\psi}$. This can be checked with the explicit form of $\ket{\bar{\psi}}$ provided in \cite{Sch15}; here we give a neat proof for symmetric states, using the symmetric formalism previously introduced in Sec.~\ref{sec:tensor}. Let $\ket{\psi}$ be a symmetric $N$-qubit state. Using representation \eqref{projrho}, it can be written as $\ketbra{\psi}=2^{-N} x_{\mu_1\mu_2\ldots\mu_N}^{\psi} S_{\mu_1\mu_2\ldots\mu_N}$. Then we have
\begin{equation}
    \langle \psi \ket{\bar{\psi}}=\tr \left(\ketbra{\psi}\fn^{\otimes N}\right) =\frac{1}{2^N} x_{\mu_1\mu_2\ldots\mu_N}^{\psi} \prod_{i=1}^N \tr \left(\fn \sigma_{\mu_i} \right).
\end{equation}
The second equality is obtained using Eq.~\eqref{matrixS} and the fact that the symmetric operator $\fn^{\otimes N}$ commutes with the projector (\ref{projps}) onto the symmetric subspace $\cals$. Since $\tr({\fn})=0$ and $\fn\sigma_j+\sigma_j\fn=0$ for all $j$, we have that $\tr \left(\fn \sigma_{\mu} \right)=0$ for all $\mu$, which completes the proof.

The antistates introduced previously exhibit an elegant geometric interpretation in the Majorana representation. This representation allows to visualize a pure symmetric state of $N$ qubits as a set of $N$ points on the unit sphere. For one qubit, this coincides with the usual Bloch representation. For several qubits, each qubit is associated to its Bloch point and symmetry allows to put all of them on the same sphere. In this picture, fully separable symmetric states correspond to $N$ degenerate points. The points of the Majorana representation of the antistate $\ket{\bar{\psi}}$ are diametrically opposite to those of the initial state $\ket{\psi}$. This is the generalization of the one-qubit case discussed in~\cite{Buz99}.

\subsection{Spectral properties of $\rho\in\nocorset_N$}%symmetric states with no $N$-partite correlations}
\label{secspec}
Let $\rho\in\nocorset_N$ and let $\ket{\psi}$ be a normalized eigenstate of $\rho$ with eigenvalue $\lambda$. The operator $\fn$ has the remarkable property that its $N$-fold tensor product commutes with $\rho$. Indeed, thanks to the anticommutation property $\fn\sigma_j+\sigma_j\fn=0$ for all $j$, we have
\begin{equation}\label{Nsr}
    \fn^{\otimes N}S_{\mu_1\mu_2\ldots\mu_N}=(-1)^{c(\mu_1,\ldots,\mu_N)}S_{\mu_1\mu_2\ldots\mu_N}  \fn^{\otimes N},
\end{equation}
where $c(\mu_1,\ldots,\mu_N)$ is the number of nonzero indices $\mu_i$. Using Eq.~(\ref{matrixS}) together with the fact that $\fn^{\otimes N}$ commutes with $P_{\cals}$ and expressing $\rho$ in the representation \eqref{projrho} as $\rho=2^{-N} x_{\mu_1\mu_2\ldots\mu_N} S_{\mu_1\mu_2\ldots\mu_N}$, only coefficients with an even number of nonzero indices appear in the expansion because of the property \eqref{nocorrsym}. Using (\ref{Nsr}), we get $\fn^{\otimes N}\rho=\rho \fn^{\otimes N}$.

As a consequence, $\rho \ket{\bar{\psi}} = \rho\fn^{\otimes N}\ket{\psi} = \fn^{\otimes N} \rho \ket{\psi}=\lambda^* \fn^{\otimes N} \ket{\psi}=\lambda \ket{\bar{\psi}}$. Thus $\ket{\bar{\psi}}$ is also an eigenstate of $\rho$ with eigenvalue $\lambda$ which is orthogonal to $\ket{\psi}$. If $\rho$ has an other eigenstate $\ket{\phi}$ with same eigenvalue $\lambda$ ($\ket{\phi}$ can be taken normalized and orthogonal to $\ket{\psi}$ and $\ket{\bar{\psi}}$ without loss of generality), the antistate $\ket{\bar{\phi}}$ will also be orthogonal to $\ket{\psi}$ and $\ket{\bar{\psi}}$ because $\fn$ is antiunitary. Repeating this procedure, we can find an orthonormal basis of the eigenspace of $\rho$ for the eigenvalue $\lambda$ containing pairs of states and antistates. Using this construction on all its eigenspaces, $\rho$ can be written as
\begin{equation}
\label{decomprho}
    \rho=\sum_{i=0}^M\lambda_i\left(\ket{\psi_i}\bra{\psi_i}+\ket{\bar{\psi}_i}\bra{\bar{\psi}_i}\right),
\end{equation}
where $\ket{\psi_i}$ and $\ket{\bar{\psi}_i}$ are eigenstates of $\rho$ with eigenvalue $\lambda_i$, and $\sum_i\lambda_i=1/2$. Equation~(\ref{decomprho}) implies that the eigenvalues of a state $\rho\in\nocorset_N$ have an even degeneracy. As a consequence, the purity of $\rho$ has a lower upper bound than the usual bound $\tr\rho^2\leq 1$, specifically
\begin{equation}
\label{ineqpurity}
\tr\rho^2\leq\frac{1}{2}.
\end{equation}
Indeed, the largest purity is reached when all $\lambda_i$ but one are zero. The double degeneracy and the normalization of the state imply that there are two nonzero eigenvalues, which are both equal to $1/2$, leading to a maximal purity $\tr\rho^2=1/2$. Since any state of the form (\ref{decomprho}) is a state with no $N$-partite correlations (as was already shown in \cite{Sch15}), this form provides a characterization of elements of $\nocorset_N$.

%%%%%%%%%%%%%%%%%%%%%%%%%%%%%%%%%%%%%%%%%%%%%%%%%%%%%%%%%%%%%%%%%%%%%%%%%
%%%%%%%%%%%%%%%%%%%%%%%%%%%%%%%%%%%%%%%%%%%%%%%%%%%%%%%%%%%%%%%%%%%%%%%%%
\section{Entanglement criteria} \label{sec:entcrit}
%%%%%%%%%%%%%%%%%%%%%%%%%%%%%%%%%%%%%%%%%%%%%%%%%%%%%%%%%%%%%%%%%%%%%%%%%
%%%%%%%%%%%%%%%%%%%%%%%%%%%%%%%%%%%%%%%%%%%%%%%%%%%%%%%%%%%%%%%%%%%%%%%%%

%%%%%%%%%%%%%%%%%%%%%%%%%%%%%%%%%%%%%%%%%%%%%%%%%%%%%%%%%%%%%%%%%%%%%%%%%
\subsection{A sufficient entanglement criterion}
%%%%%%%%%%%%%%%%%%%%%%%%%%%%%%%%%%%%%%%%%%%%%%%%%%%%%%%%%%%%%%%%%%%%%%%%%

A sufficient criterion for genuine entanglement was obtained in \cite{Bad08}. Following this approach we now derive a sufficient criterion for genuine entanglement of symmetric states. Let $S^2$ denote the unit sphere in $\bbr^3$, and $\ket{n}$ be the fully separable state (\ref{symsep}) associated with ${\bf n}\in S^2$. If $\rho$ is a symmetric state such that 
\begin{equation}
\label{genuine}
\forall \: {\bf n}\in S^2,\quad \bra{n}\rho\ket{n}<\tr \rho^2,
\end{equation}
then $\rho$ is genuinely entangled.

Indeed, suppose $\rho$ is not genuinely entangled. Then it is fully separable and therefore can be written as a mixture of fully separable pure states $\ket{n^{(i)}}$, namely
\begin{equation}
\rho=\sum_ip_i\ketbra{n^{(i)}},
\end{equation}
with $0<p_i\leq 1$ and $\sum_i p_i=1$. With these notations, we have
\begin{equation}
\begin{aligned}
\label{trrho2}
\tr\rho^2&=\sum_i p_i\bra{n^{(i)}}\rho\ket{n^{(i)}}\\
&\leqslant\max_i\;\bra{n^{(i)}}\rho\ket{n^{(i)}},
\end{aligned}
\end{equation}
since $p_i$ are positive and sum up to 1. The state $\ket{n^{(i)}}$ achieving the maximum in \eqref{trrho2} violates Eq.~\eqref{genuine}, hence the result.

In the case where $\rho$ is of rank 2, this condition is in fact necessary and sufficient, as will be shown in Section \ref{rank2sec}.

%%%%%%%%%%%%%%%%%%%%%%%%%%%%%%%%%%%%%%%%%%%%%%%%%%%%%%%%%%%%%%%%%%%%%%%%%
\subsection{A necessary and sufficient criterion in the three-qubit case}\label{subsec:3qubit}
%%%%%%%%%%%%%%%%%%%%%%%%%%%%%%%%%%%%%%%%%%%%%%%%%%%%%%%%%%%%%%%%%%%%%%%%%

As we ruled out even values of $N$, the simplest nontrivial case is $N=3$. As we show below, the set of three-qubit genuinely entangled symmetric states with no three-partite correlations can be fully characterized. For $\rho\in\nocorset_3$, Eq.~(\ref{projrho}) reduces to
\begin{equation}\label{rho3}
\rho=\frac{1}{8}\mathbb{1}_4+\frac{3}{8}\sum_{a,b=1}^3 A_{ab} S_{ab0},
\end{equation}
with $\mathbb{1}_4$ the $4\times 4$ identity matrix, $S_{ab0}$ the matrices defined in Eq.~(\ref{matrixS}) and explicitly given by
\begin{equation}\label{eq:Sab}
S_{ab0}=\frac{J_a J_b + J_b J_a}{3} - \frac{\delta_{ab}}{2}\mathbb{1}_4,
\end{equation}
with $J_a$ the $4\times 4$ angular momentum matrices, and $A$ the $3\times3$ real symmetric matrix $(x_{ab0})_{1\leq a ,b\leq 3}$. Equation (\ref{traceless}) implies $\tr A=1$. The possible values of $A_{ab}$ are further constrained by the fact that $\rho$ has to be a positive semidefinite matrix. The characteristic polynomial of $\rho$ can be put under the form
\begin{equation}
\det(z\mathbb{1}-\rho)=\left(z^2-\frac12 z+\frac{3b}{16}\right)^2,
\end{equation}
with $b=(1-\tr A^2)/2$. This readily implies that the eigenvalues of $\rho$ are nonnegative if and only if $b\geq 0$, that is, $\tr A^2\leq 1$.

For a state $\rho$ acting on a Hilbert space $\mathcal{H}_1\otimes\mathcal{H}_2$ with $\mathcal{H}_1$ isomorphic to $\mathbb{C}^2$ and $\mathcal{H}_2$ isomorphic to $\mathbb{C}^3$, a necessary and sufficient condition of separability is the celebrated Peres-Horodecki criterion~\cite{Hor96}. This criterion states that the partial transpose of $\rho$ is positive semidefinite if and only if $\rho$ is separable. Since any fully symmetric three-qubit state can be expressed in the canonical basis of $\mathbb{C}^{2}\otimes \mathbb{C}^{3}$, we can apply this criterion to $\rho\in\nocorset_3$~(see e.g.~\cite{Aug12}). Here, we will denote by $\rho^{\mathrm{PT}}$ the partial transpose performed on the first qubit (since we are dealing with symmetric states, the qubit on which the partial transpose is performed does not matter). In~\cite{Boh16}, it was shown that $\rho^{\mathrm{PT}}$ is positive semidefinite if and only if an Hermitian $8\times 8$ matrix $T$ given in terms of $x_{\mu_1\mu_2\mu_3}$ is positive semidefinite (the explicit expression for $T$ can be found at Eq.~(44) of Ref.~\cite{Boh16}). From this explicit form and using the relations (\ref{nocorrsym}), it can be checked that the characteristic polynomial of $T$ can be expressed as $\det(z\mathbb{1}_8-T)= z^2 q(z)^2$ with 
\begin{equation}\label{charpolT}
\begin{aligned}
q(z)={}&z^3-2z^2+3\frac{(\tr A)^2-\tr A^2}{2}z\\
&  -2\frac{(\tr A)^3-3(\tr A)^2\tr A^2+2\tr A^3}{3}.
\end{aligned}
\end{equation}
%in the case where conditions (\ref{nocorrsym}) are fulfilled.
The matrix $T$ is positive semidefinite if and only if all roots of $q(z)$ are nonnegative. Moreover, the characteristic polynomial of $A$ is $\det(z\mathbb{1}_3-A)$ and reads
\begin{equation}\label{charpolA}
\begin{aligned}
&z^3-z^2+\frac{(\tr A)^2-\tr A^2}{2}z\\
&-\frac{(\tr A)^3-3(\tr A)^2\tr A^2+2\tr A^3}{6}.
\end{aligned}
\end{equation}
From Eqs.~(\ref{charpolT}) and (\ref{charpolA}), it appears that the roots of $q(z)$ are all nonnegative if and only if the roots of $A$ are all nonnegative (this follows from Descartes' rule of signs). Thus, $\rho^{\mathrm{PT}}$ is positive semidefinite if and only if $A$ is positive semidefinite. This provides a necessary and sufficient separability criterion in terms of the matrix $A$, namely $A\geq 0$. Using \eqref{correl}, we can see $A$ as the two-partite correlation matrix $A=(\langle \sigma_{ab}\rangle)_{1\leq a,b\leq 3}$, with $\sigma_{ab}=\sigma_a\otimes\sigma_b\otimes \mathbb{1}_2$ and $\langle \sigma_{ab}\rangle_\rho=\tr (\rho\,\sigma_{ab})$. The necessary and sufficient separability criterion can then be reformulated, for any $\rho\in\nocorset_3$, as 
\begin{equation}\label{rhosepApos3}
%\begin{aligned}
%&\forall~\rho\in\nocorset_3,\\[4pt]
\rho~\mathrm{separable}\;\Leftrightarrow\; \begin{pmatrix}
\langle \sigma_{11}\rangle_\rho & \langle \sigma_{12}\rangle_\rho & \langle \sigma_{13}\rangle_\rho \\ 
\langle \sigma_{21}\rangle_\rho & \langle \sigma_{22}\rangle_\rho & \langle \sigma_{23}\rangle_\rho \\ 
\langle \sigma_{31}\rangle_\rho & \langle \sigma_{32}\rangle_\rho & \langle \sigma_{33}\rangle_\rho
\end{pmatrix} \geq 0.
\end{equation}

Let us now consider the two-qubit reduced density matrix $\rho_{2}$ of $\rho\in\nocorset_3$. According to Eq.~(\ref{projrhok}), the tensor coordinates of $\rho_{2}$ are the $x_{\mu_1\mu_20}$. From Ref.~\cite{Boh16}, the partial transpose $\rho_2^{\mathrm{PT}}$ is positive semidefinite if and only if the $4\times 4$ matrix 
$(x_{\mu_1\mu_20})_{0\leq\mu_1,\mu_2\leq 3}$ is positive semidefinite. This latter matrix is block-diagonal, with upper left $1\times1$ block being the identity and the bottom right $3\times3$ block given by matrix $A$. Therefore positivity of $(x_{\mu_1\mu_20})_{0\leq\mu_1,\mu_2\leq 3}$ is equivalent to positivity of $A$. These equivalences together with the result \eqref{rhosepApos3} show that for any $\rho\in\nocorset_3$, 
\begin{equation}
\rho~\mathrm{separable}\;\Leftrightarrow\;\rho_2~\mathrm{separable}.
\end{equation}
In physical terms, this means that one cannot end up in a separable state by tracing out one qubit from an entangled three-qubit symmetric state with no three-partite correlations. Therefore, the entanglement in such states has some robustness.

%%%%%%%%%%%%%%%%%%%%%%%%%%%%%%%%%%%%%%%%%%%%%%%%%%%%%%%%%%%%%%%%%%%%%%%%%
\subsection{Other sufficient entanglement criteria} 
%%%%%%%%%%%%%%%%%%%%%%%%%%%%%%%%%%%%%%%%%%%%%%%%%%%%%%%%%%%%%%%%%%%%%%%%%

The above criterion for three-qubit states also provides us with a sufficient entanglement criterion in the general $N$-qubit case. Indeed, if $\rho\in\nocorset_N$ is separable, then its three-qubit reduced density matrix $\rho_3$ is separable. Using (\ref{rhosepApos3}) and (\ref{projrhok}), we get a sufficient entanglement condition for $\rho$: if the $3\times 3$ matrix $A=(\langle \sigma_{ab}\rangle_\rho)_{1\leq a,b\leq 3}$, with $\sigma_{ab}=\sigma_a\otimes\sigma_b\otimes \mathbb{1}_2\otimes \cdots\otimes\mathbb{1}_2$, is not positive semidefinite, then $\rho$ is genuinely entangled. This is reminiscent of the entanglement criteria obtained for pure states in \cite{Mar13} based on two-point correlations.

It is in fact possible to obtain many more sufficient entanglement criteria from the PPT criteria applied to $\rho$ or to its $k$-qubit reduced density matrices. As shown in~\cite{Boh16}, the partial transpose matrices, and their positivity, can be expressed in terms of the $x_{\mu_1\ldots\mu_N}$ in a simple way. As we saw above, the partial transpose $\rho_2^{\mathrm{PT}}$ can be related with the $4\times 4$ matrix 
$(x_{\mu_1\mu_20})_{0\leq\mu_1,\mu_2\leq 3}$ and positivity of $\rho_2^{\mathrm{PT}}$ is equivalent to the right-hand side of (\ref{rhosepApos3}). Similarly, positivity of the partial transpose $\rho_4^{\mathrm{PT}}$ is equivalent to positivity of the $16\times 16$ matrix indexed by the 16 pairs $(\mu_1,\mu_2)$ and $(\mu_3,\mu_4)$ and whose entries are the $x_{\mu_1\mu_2 \mu_3 \mu_4 0}$. All criteria that can be obtained in the same way lead to sufficient conditions of genuine entanglement in terms of correlators. However, these criteria lead to conditions that involve polynomials of high degree in $x_{\mu_1\ldots\mu_N}$ (the simple criterion $A\geq 0$ yields a polynomial of degree 3 in the $x_{ab0}$). In contrast, the criterion (\ref{genuine}) is of degree 2, and thus easier to deal with. In Sec.~\ref{sec:families}, our construction will be based on criterion (\ref{genuine}).

%%%%%%%%%%%%%%%%%%%%%%%%%%%%%%%%%%%%%%%%%%%%%%%%%%%%%%%%%%%%%%%%%%%%%%%%%
%%%%%%%%%%%%%%%%%%%%%%%%%%%%%%%%%%%%%%%%%%%%%%%%%%%%%%%%%%%%%%%%%%%%%%%%%
\section{Application to three-qubit states} \label{sec:threequbits}
%%%%%%%%%%%%%%%%%%%%%%%%%%%%%%%%%%%%%%%%%%%%%%%%%%%%%%%%%%%%%%%%%%%%%%%%%
%%%%%%%%%%%%%%%%%%%%%%%%%%%%%%%%%%%%%%%%%%%%%%%%%%%%%%%%%%%%%%%%%%%%%%%%%

\begin{figure}
    \centering
    \includegraphics*[width=0.8\linewidth]{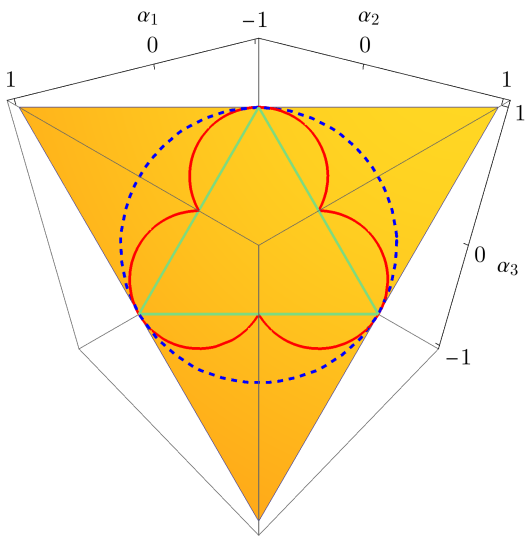}
    \caption{Three-qubit symmetric states in the space of eigenvalues $\alpha_i$ of $A$. The orange plane (triangle down) corresponds to the condition $\tr A=\sum_i\alpha_i=1$. Points inside the dashed blue circle defined by $\tr A^2=\sum_i\alpha_i^2=1$ correspond to physical states $\rho\geq 0$. Points inside the green triangle up, defined by $\alpha_i\geq 0$, correspond to separable states. Points outside the green triangle up and inside the dashed blue circle correspond to genuinely entangled states. Points lying between the dashed circle and the solid red trilobe defined by $\max_i\alpha_i=\sum_i\alpha_i^2$ correspond to genuinely entangled states detected by the sufficient criterion (\ref{genuine}).}
    \label{fig1}
\end{figure}

Although we obtained a necessary and sufficient condition for three qubits in Sec.~\ref{sec:entcrit}, it is instructive to apply our general approach based on criterion (\ref{genuine}) to this simple case.
Condition \eqref{nocorrsym} implies that for a state $\rho$ in the representation \eqref{projrho} one has $x_{abc}=0$ and $x_{a00}=0$ for $a,b,c=1,2,3$. The only nonzero coordinates $x_{\mu_1\mu_2\mu_3}$ are the $x_{ab0}$ for $1\leq a,b\leq 3$, and $x_{000}=1$, leading to the representation (\ref{rho3}). We label by $\alpha_i$ the eigenvalues of $A$. One easily calculates $\bra{n}\rho\ket{n}=(1+3\,{\bf n}^TA{\bf n})/8$ and $\tr\rho^2=(1+3\,\tr A^2)/8$, so that condition \eqref{genuine} becomes
\begin{equation}
\forall \: {\bf n}\in S^2,\qquad {\bf n}^TA{\bf n}<\tr A^2.
\end{equation}
In terms of the eigenvalues of $A$, this condition can be reexpressed as
\begin{equation}
\label{condN3}
\max_i\alpha_i<\sum_i\alpha_i^2.
\end{equation}
As discussed above, since $\rho$ has to be a semidefinite positive matrix, $A$ must be such that $\tr A^2\leq 1$. Taking into account the fact that $\tr A=1$, the $\alpha_i$ must fulfill the two additional constraints
\begin{equation}
\label{addconstr}
\sum_i\alpha_i^2\leq 1\quad \textrm{and}\quad\sum_i\alpha_i=1.
\end{equation}
Among the set of triplets $(\alpha_1,\alpha_2,\alpha_3)\in\bbr^3$ fulfilling the conditions \eqref{addconstr}, one can easily find those verifying condition \eqref{condN3}. They are depicted in Fig.~\ref{fig1}. Equation (\ref{addconstr}) imposes that they are restricted to the plane $\sum_i\alpha_i=1$ and to the interior of the sphere $\sum_i\alpha_i^2\leq 1$ (see Fig.~\ref{fig1}). The equality $\alpha_1=\sum_i\alpha_i^2$ is the equation of a sphere of radius $1/2$ and center $(1/2,0,0)$, whose intersection with the plane is a circle. A similar analysis for $\alpha_2$ and $\alpha_3$ implies that the solutions to \eqref{condN3} lie strictly outside the three solid red circles of Fig.~\ref{fig1}. Finally, the triplets of solutions of \eqref{condN3} and \eqref{addconstr} lie in the region between the dashed blue circle and the solid red trilobe in Fig.~\ref{fig1}. These points correspond to the genuinely entangled states detected by the sufficient criterion (\ref{genuine}).\\

For $\rho\in\nocorset_3$, the criterion \eqref{rhosepApos3} gives us a necessary and sufficient entanglement criterion in terms of the matrix $A$ defined above. Thus $\rho$ is separable if and and only if all $\alpha_i$ are nonnegative. In Fig.~\ref{fig1}, the corresponding region is the green triangle up and its interior. Hence $\rho$ is genuinely entangled if and only if its associated point lies outside the triangle but inside the dashed blue circle. Points outside the triangle but inside the trilobe are associated with genuinely entangled states not detected by criterion (\ref{genuine}). 

%%%%%%%%%%%%%%%%%%%%%%%%%%%%%%%%%%%%%%%%%%%%%%%%%%%%%%%%%%%%%%%%%%%%%%%%%
%%%%%%%%%%%%%%%%%%%%%%%%%%%%%%%%%%%%%%%%%%%%%%%%%%%%%%%%%%%%%%%%%%%%%%%%%
\section{Construction of families} \label{sec:families}
%%%%%%%%%%%%%%%%%%%%%%%%%%%%%%%%%%%%%%%%%%%%%%%%%%%%%%%%%%%%%%%%%%%%%%%%%
%%%%%%%%%%%%%%%%%%%%%%%%%%%%%%%%%%%%%%%%%%%%%%%%%%%%%%%%%%%%%%%%%%%%%%%%%

%%%%%%%%%%%%%%%%%%%%%%%%%%%%%%%%%%%%%%%%%%%%%%%%%%%%%%%%%%%%%%%%%%%%%%%%%
\subsection{Rank-2 density matrices}\label{rank2sec}
%%%%%%%%%%%%%%%%%%%%%%%%%%%%%%%%%%%%%%%%%%%%%%%%%%%%%%%%%%%%%%%%%%%%%%%%%

\subsubsection{A necessary and sufficient entanglement criterion}
\label{rank2abs}
Let $\rho\in\nocorset_N$ be of rank 2; it has two nonzero eigenvalues, which have to be equal according to the results of section \ref{antistates}. Since $\tr\rho=1$, both eigenvalues are equal to $1/2$, so that there exists a pure state $\ket{\psi}$ such that
\begin{equation}
\label{rhorank2}
\rho=\frac12\ketbra{\psi}+\frac12\ketbra{\bar{\psi}}.
\end{equation}
Any fully separable pure symmetric state $\ket{n}$ can be decomposed as
\begin{equation}
\ket{n}=\langle \psi|n\rangle \ket{\psi}+\langle\bar{\psi}|n\rangle \ket{\bar{\psi}}+\ket{\phi},
\end{equation}
where $\ket{\phi}$ is orthogonal to both $\ket{\psi}$ and $\ket{\bar{\psi}}$ (we recall that the latter two states are orthogonal). The overlap $\bra{n}\rho\ket{n}$ then reads
\begin{equation}
\label{nrhon}
\bra{n}\rho\ket{n}=\frac12|\langle \psi|n\rangle|^2+\frac12|\langle\bar{\psi}|n\rangle |^2=\frac12\left(1-\langle \phi|\phi\rangle\right).
\end{equation}
Since for the state $\rho$ given by Eq.~(\ref{rhorank2}), one has $\tr\rho^2=1/2$, Eq.~(\ref{nrhon}) yields
\begin{equation}\label{rank2cond}
\bra{n}\rho\ket{n}\leq \frac{1}{2}=\tr\rho^2.
\end{equation}
Thus, inequality or equality is always achieved in \eqref{genuine}. In fact, Eq.~(\ref{genuine}) is a necessary and sufficient condition for genuine entanglement in the case of rank-2 density matrices. Indeed, suppose Eq.~(\ref{genuine}) is violated for some $\ket{n}$. Then, from Eq.~(\ref{rank2cond}) one must have $\bra{n}\rho\ket{n} = \tr\rho^2=1/2$ which using \eqref{nrhon} implies that $\ket{\phi}=0$, so that $\ket{n}$ lies in the subspace spanned by $\ket{\psi}$ and $\ket{\bar{\psi}}$, which is the eigenspace of $\rho$ associated with eigenvalue $1/2$. In particular one must have the decomposition
\begin{equation}
\label{rhorank2beta}
\rho=\frac12\ketbra{n}+\frac12\ketbra{\bar{n}}.
\end{equation}
Therefore, $\rho$ is a mixture of two fully separable states, thus $\rho$ is separable. Hence, if $\rho$ is genuinely entangled, then Eq.~(\ref{genuine}) must hold. Thus, a rank-2 state $\rho\in\nocorset_N$ is genuinely entangled if and only if
\begin{equation}
\label{genuinerank2}
\forall\: {\bf n}\in S^2,\quad \bra{n}\rho\ket{n}<\frac{1}{2}.
\end{equation}
It is separable if and only if there exists $\ket{n}$ such that $\bra{n}\rho\ket{n} = \tr\rho^2=1/2$, which is equivalent to \eqref{rhorank2beta}.

\subsubsection{A necessary separability criterion}

Let $\rho$ be a rank-2 symmetric state of the form \eqref{rhorank2}, thus with no $N$-partite correlations. Expressing \eqref{rhorank2} in terms of the coordinates $x^\rho_{\mu_1\mu_2\ldots\mu_N}$ of $\rho$ and the coordinates $x^\psi_{\mu_1\mu_2\ldots\mu_N}$ of $\ketbra{\psi}$ in the expansion (\ref{projrho}), we get
\begin{equation}
x^\rho_{\mu_1\mu_2\ldots\mu_N}=\frac{1+(-1)^{c(\mu_1,\ldots,\mu_N)}}{2}x^\psi_{\mu_1\mu_2\ldots\mu_N},
\end{equation}
where $c(\mu_1,\ldots,\mu_N)$ is the number of nonzero indices $\mu_i$. According to the previous subsection, it is separable if and only if it can be written as in~\eqref{rhorank2beta}. In terms of tensor coordinates, it is equivalent to
\begin{equation}
x^\rho_{\mu_1\mu_2\ldots\mu_N}=\frac{1+(-1)^{c(\mu_1,\ldots,\mu_N)}}{2}n_{\mu_1}\ldots n_{\mu_N}.
\end{equation}
This implies that for an even number of nonzero indices, we have $x_{\mu_1\ldots\mu_n}^{\rho}=n_{\mu_1}\ldots n_{\mu_N}=x^\psi_{\mu_1\mu_2\ldots\mu_N}$. In particular, if all but two indices are zero, we have $x^\rho_{ab 0\ldots 0}=n_an_b$, so that the matrix $A=(x^\rho_{ab 0\ldots 0})_{1\leq a,b\leq 3}=(x^\psi_{ab 0\ldots 0})_{1\leq a,b\leq 3}=(\langle \sigma_{ab}\rangle_\rho)_{1\leq a,b\leq 3}$ is of rank one, where $\sigma_{ab}=\sigma_a\otimes\sigma_b\otimes \mathbb{1}_2\otimes \cdots\otimes\mathbb{1}_2$. 

We therefore get the following necessary condition for separability of rank-2 states,
\begin{equation}\label{sepcrit}
\rho~\mathrm{separable}\quad\Rightarrow\quad \mathrm{rank}A=1.
\end{equation}

\subsubsection{Explicit examples}
The above considerations allow us to construct families of genuinely entangled states of $\nocorset_N$. Indeed, for any state $\ket{\psi}$, the mixed state $\rho=(\ket{\psi}\bra{\psi}+\ket{\bar{\psi}}\bra{\bar{\psi}})/2$ has no $N$-partite correlations. Choosing $\ket{\psi}$ in such a way that $A=(x^\psi_{ab 0\ldots 0})_{1\leq a,b\leq 3}$ is not of rank one warrants that $\rho$ is also genuinely entangled. 

This construction can be achieved for instance for $\ket{\psi}=(\ket{D_{N}^{(r)}}+\ket{D_{N}^{(N-r)}})/\sqrt{2}$, which is a superposition of two Dicke states defined in \eqref{DickeStates}. Such states have already been studied in \cite{Tra17}. Here they provide a simple illustration of the rank-2 entanglement criterion derived in Section \ref{rank2abs}. The coefficients of the matrix $A^{\psi}$ are $x_{ab0\ldots0}^{\psi}=\bra{\psi}\sigma_{ab}\ket{\psi}$, so that 
\begin{equation}
\begin{aligned}
    x_{ab0\ldots0}^{\psi}=\frac{1}{2} & \left(\bra{D_{N}^{(r)}}\sigma_{ab}\ket{D_{N}^{(r)}}\right.+\bra{D_{N}^{(N-r)}}\sigma_{ab}\ket{D_{N}^{(N-r)}}\\[0.2cm]
    &+\bra{D_{N}^{(r)}}\sigma_{ab}\ket{D_{N}^{(N-r)}}\left.+\bra{D_{N}^{(N-r)}}\sigma_{ab}\ket{D_{N}^{(r)}}\right).
\end{aligned}
\label{xabrk2}
\end{equation}
The last two terms of Eq.~(\ref{xabrk2}) can be shown to vanish for odd $N$ through an argument on the parity of the number of excitations \cite{Tra17}. In order to apply $\sigma_{ab}$ on the Dicke states, we decompose them as 
\begin{multline}
\ket{D_{N}^{(r)}}= \sqrt{\frac{\binom{N-2}{r}}{\binom{N}{r}}}\ket{00}\ket{D_{N-2}^{(r)}}
+\sqrt{\frac{\binom{N-2}{r-2}}{\binom{N}{r}}}\ket{11}\ket{D_{N-2}^{(r-2)}}\\
+\sqrt{\frac{\binom{N-2}{r-1}}{\binom{N}{r}}}(\ket{01}+\ket{10})\ket{D_{N-2}^{(r-1)}}.
\end{multline} 
Equation \eqref{xabrk2} then reduces to
\begin{multline}\label{xab}
x_{ab0\ldots0}^{\psi}=\frac{2r(N-r)}{N(N-1)}(\delta_{a,1}\delta_{b,1}+\delta_{a,2}\delta_{b,2})\\+\frac{(N-2r)^2-N}{N(N-1)}\delta_{a,3}\delta_{b,3}.
\end{multline}
The rank of $A^{\psi}$ is then equal to one if $r=0$ or $N$, to two if $r=(N-\sqrt{N})/2$, and to three otherwise. Thus, any mixed state defined by \eqref{rhorank2} with $\ket{\psi}=(\ket{D_{N}^{(r)}}+\ket{D_{N}^{(N-r)}})/\sqrt{2}$ and $1\leq r\leq N-1$ is a genuinely entangled state of $\nocorset_N$.

%%%%%%%%%%%%%%%%%%%%%%%%%%%%%%%%%%%%%%%%%%%%%%%%%%%%%%%%%%%%%%%%%%%%%%%%%
\subsection{Arbitrary rank}
%%%%%%%%%%%%%%%%%%%%%%%%%%%%%%%%%%%%%%%%%%%%%%%%%%%%%%%%%%%%%%%%%%%%%%%%%

\subsubsection{General construction}
The above construction can be generalized to any even rank (as shown in Sec.~\ref{secspec}, any genuinely entangled $N$-qubit state with no $N$-partite correlation has its eigenvalues always twice degenerate, so that no odd-rank example can exist). From the decomposition \eqref{decomprho} of any state in $\nocorset_N$, the sufficient genuine entanglement criterion \eqref{genuine} can be expressed as 
\begin{equation}
\forall \:{\bf n}\in S^2,\quad \sum_{i=0}^{M}\lambda_i\left(\left|\bra{\psi_i}n\rangle\right|^2+\left|\bra{\bar{\psi}_i}n\rangle\right|^2\right) < 2\sum_{i=0}^M\lambda_i^2,
\end{equation}
with $\ket{n}$ a pure symmetric fully separable state. This criterion can be rewritten as
\begin{equation}\label{sphere}
\forall {\bf n}\in S^2,\quad \sum_{i=0}^{M}(\lambda_i-c_i(\bn))^2>\sum_{i=0}^Mc_i(\bn)^2,
\end{equation}
where $c_i(\bn)=(|\bra{\psi_i}n\rangle|^2+|\bra{\bar{\psi}_i}n\rangle|^2)/4$. Thus, if for all $\bn$, the vector $(\lambda_0,\ldots,\lambda_M)$ is outside the sphere $S(\bn)$ centered at $C(\bn)=(c_0(\bn),\ldots,c_M(\bn))$ and going through the origin, then the state \eqref{decomprho} is genuinely entangled.

Let us explain how to construct arbitrary-rank examples. It suffices that one of the $\ket{\psi_i}$ in \eqref{decomprho}, for example $\ket{\psi_{i_0}}$, be such that $A=(x^{\psi_{i_0}}_{ab 0\ldots 0})_{1\leq a,b\leq 3}$ is of rank two or three, so that $\rho_{i_0}=\left(\ket{\psi_{i_0}}\bra{\psi_{i_0}}+\ket{\bar{\psi}_{i_0}}\bra{\bar{\psi}_{i_0}}\right)/2$ is genuinely entangled according to Eq.~\eqref{sepcrit}. Then, the vector $E=(0,\ldots,0,1/2,0,\ldots,0)\in \bbr^M$, where the nonzero component is the ${i_0}$th, is outside all spheres $S(\bn)$ since Eq.~(\ref{sphere}) reduces in this case to $c_{i_0}(\bn)<1/4$, which according to Eq.~(\ref{genuinerank2}) is equivalent to the fact that $\rho_{i_0}$ is genuinely entangled. Thus, for any fixed $\bn$, the distance $d(E,S(\bn))$ between $E$ and the sphere $S(\bn)$ is such that $d(E,S(\bn))>0$. Since $\bn$ is parametrized by spherical angles $\theta$ and $\varphi$ which vary in the compact set $[0,\pi]\times[0,2\pi]$, the minimum of $d(E,S(\bn))$ over all $\bn$ is reached for some $\bn_0$, so that $\inf_\bn d(E,S(\bn))=d(E,S(\bn_0))>0$. Therefore, there are vectors $(\lambda_0,\ldots,\lambda_M)$ in the vicinity of $E$ such that $\lambda_i\geq0$, $\sum_{i=0}^M\lambda_i=1/2$ and the genuine entanglement criterion \eqref{sphere} is fulfilled. This shows that once $\ket{\psi_{i_0}}$ has been chosen as explained above, any choice of $\ket{\psi_i}$ with $i\ne {i_0}$ allows to construct a state of arbitrary rank of the form \eqref{decomprho} which for some values of the weights $\lambda_i$ is guaranteed to be a genuinely entangled state of $\nocorset_N$.

\subsubsection{Explicit examples}
As an illustration, we consider the case where $\ket{\psi_i}$ in (\ref{decomprho}) are chosen as 
\begin{equation}\label{psiDicke}
\ket{\psi_i}=\frac{1}{\sqrt{2}}(\ket{D_{N}^{(i)}}+\ket{D_{N}^{(N-i)}}),\quad 0\leq i\leq M.
\end{equation}
According to Eq.~(\ref{xab}), any choice $i_0\ne 0$ yields a rank-2 state $\rho_{i_0}=\left(\ket{\psi_{i_0}}\bra{\psi_{i_0}}+\ket{\bar{\psi}_{i_0}}\bra{\bar{\psi}_{i_0}}\right)/2$ which is genuinely entangled. In the vicinity of each $E=(0,\ldots,0,1/2,0,\ldots,0)\in \bbr^M$, there exist values of $\lambda_i$ giving arbitrary rank genuinely entangled states.

\begin{figure*}
    \centering
    \begin{tabular}{cc@{\hspace{2cm}}c}
    (a) && (b) \\[4pt]
    \raisebox{-.5\height}{\includegraphics*[width=0.325\linewidth]{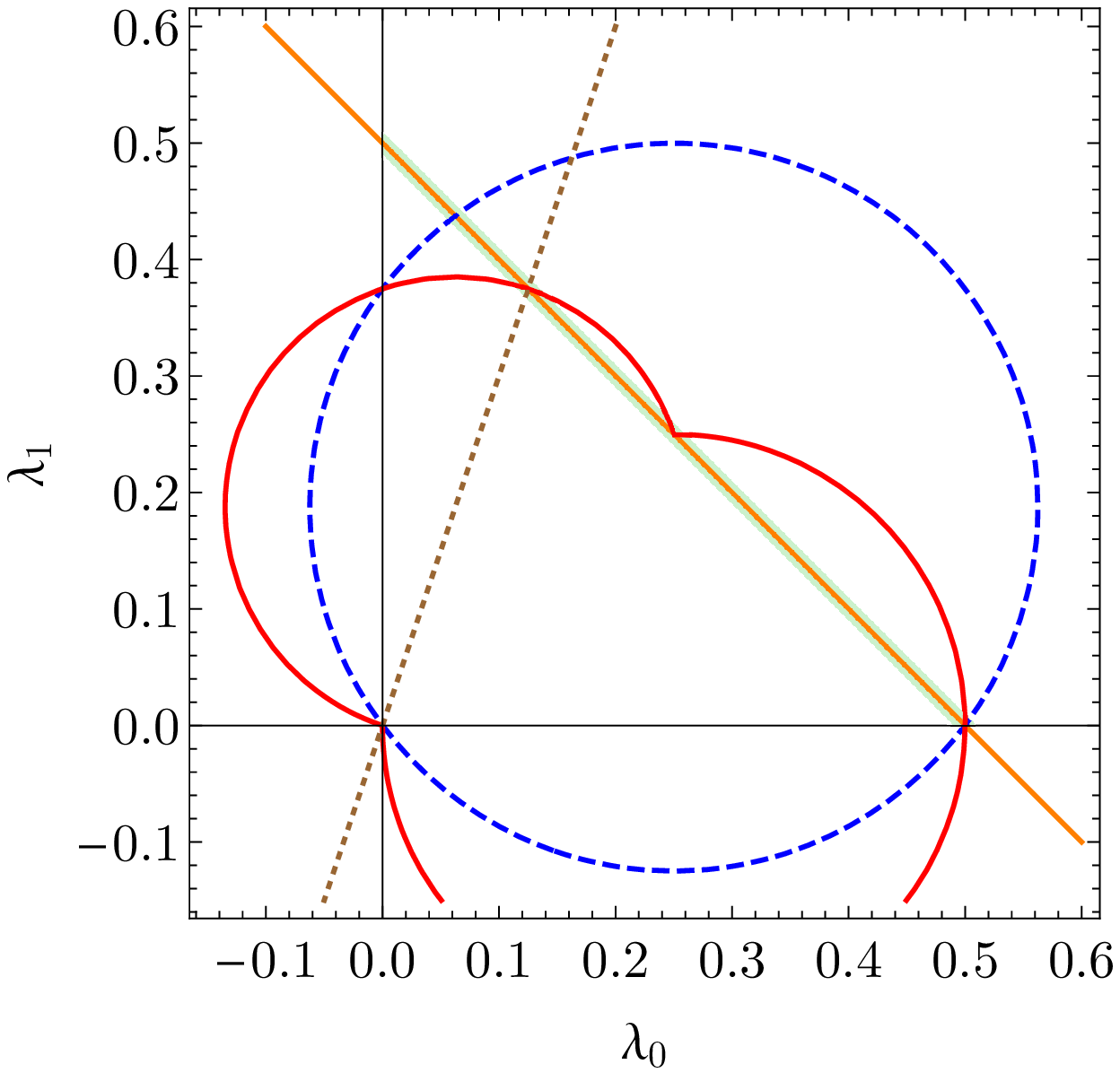}} && \raisebox{-.5\height}{\includegraphics*[width=0.35\linewidth]{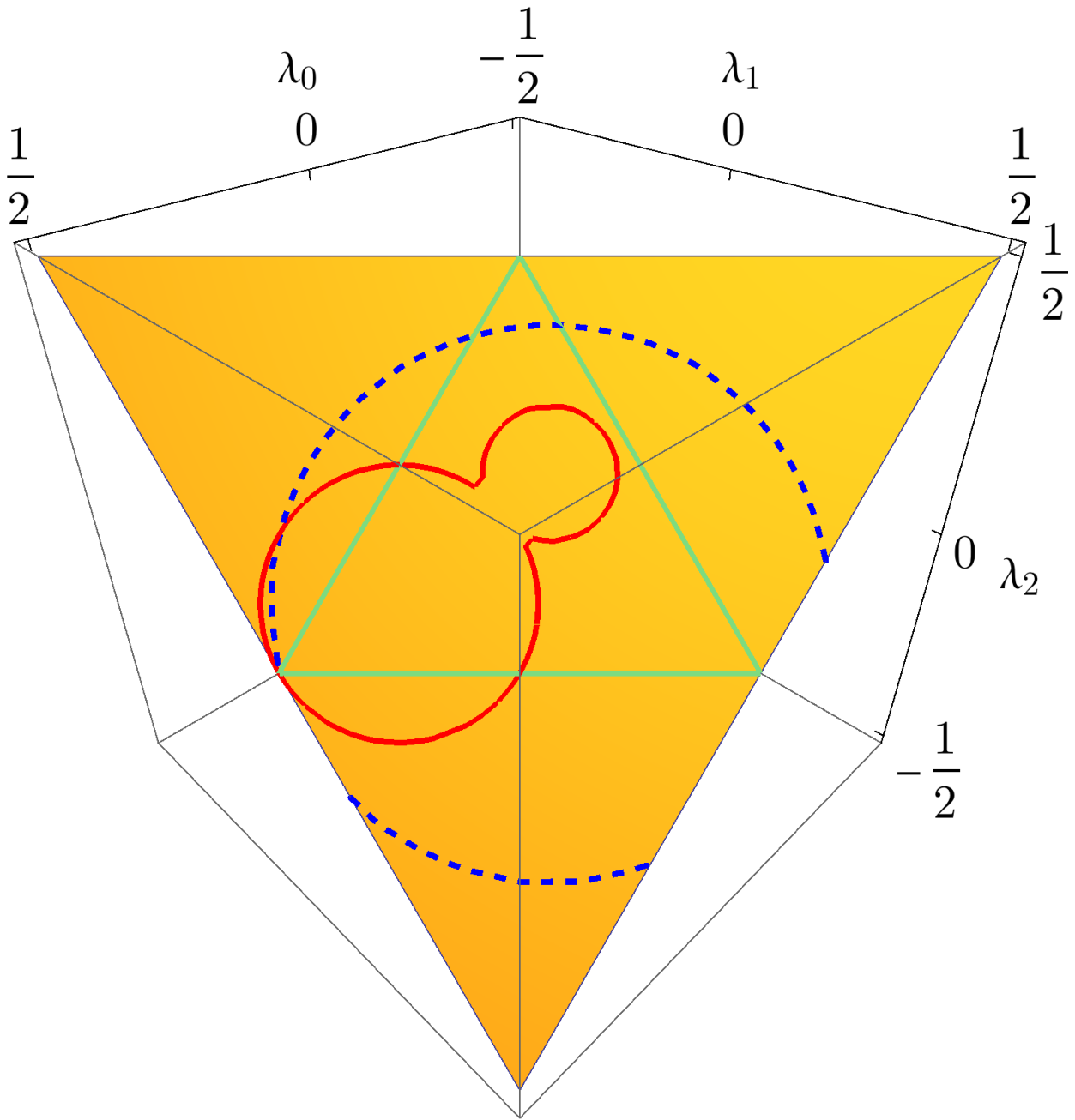}}
    \end{tabular}
    \caption{Three-qubit (a) and five-qubit (b) symmetric states in the space of eigenvalues of $\rho$. The orange line $\lambda_0+\lambda_1=1/2$ for $N=3$, and the large orange triangle down for $N=5$ correspond to the condition $\tr \rho=2\sum_i\lambda_i=1$. Points where all $\lambda_i\geq 0$ correspond to physical states $\rho\geq 0$ and are depicted by the thick green line $0\leq\lambda_0\leq 1/2$ for $N=3$, the green triangle up for $N=5$. In (a) and (b), the equality in (\ref{eqn:sphere}) is the dashed blue curve and the numerically solved criterion (\ref{genuine}) is the solid red curve. In (a), the dotted straight line is the analytical result $\lambda_1=3\lambda_0$. In (b), only the intersection with the normalization plane is depicted.}
    \label{fig2}
\end{figure*}

In order to give an explicit region in parameter space where the constructed states are genuinely entangled, we use the explicit form of the $\ket{\psi_i}$. Since the antistate of a Dicke state is $\ket{\bar{D}_N^{(k)}}=(-1)^{N-k}\ket{D_N^{(N-k)}}$, it comes that $\ket{\psi_i}\bra{\psi_i}+\ket{\bar{\psi}_i}\bra{\bar{\psi}_i}=\ket{D_N^{(i)}}\bra{D_N^{(i)}}+\ket{D_N^{(N-i)}}\bra{D_N^{(N-i)}}$, which yields
\begin{equation}
\bra{n}\rho\ket{n}=\sum_{i=0}^{M}\lambda_i\left(\big|\bra{D_N^{(i)}}n\rangle\big|^2+\big|\bra{D_N^{(N-i)}}n\rangle\big|^2\right).
\end{equation}
Using the expansion (\ref{symsep}), we get
\begin{equation}\label{trig}
\bra{n}\rho\ket{n}=\sum_{i=0}^{M}\lambda_i u_i(\theta),
\end{equation}
with $u_i(\theta)$ defined as
\begin{equation}
\binom{N}{i}\left[\frac{\sin\theta}{2}\right]^{2i}\frac{\left(1-\cos\theta\right)^{N-2i}+\left(1+\cos\theta\right)^{N-2i}}{2^{N-2i}}.
\end{equation}
One therefore has the upper bound
\begin{equation}\label{trigN}
\bra{n}\rho\ket{n}\leq \sum_{i=0}^{M}\lambda_i \,\max_{\theta}u_i(\theta).
\end{equation}
Taking $\lambda_i$ such that $\sum_i\lambda_i \,\max_{\theta}u_i(\theta)<2\sum_i\lambda_i^2$  ensures that the state $\rho$ is genuinely entangled. This inequality gives a constraint on the weight of $\rho_{i_0}$ in the mixture $\rho$ given by Eq.~(\ref{decomprho}), which cannot be too small. For arbitrary $N$, one can find a less stringent analytical bound. Since the function $f_k(x)=[(1-\cos x)^k+(1+\cos x)^k]/2^k$ takes values between 0 and 1, we have $u_i(\theta)\leq \binom{N}{i}2^{-2i}$ so that any state with $\lambda_i$ such that
\begin{equation}
    \sum_{i=0}^{M}\binom{N}{i}\frac{\lambda_i}{2^{2i}}<2\sum_{i=0}^{M}\lambda_i^2,
    \label{eqn:sphere}
\end{equation}
$\lambda_i\geq 0$ and $\sum_{i=0}^{M}\lambda_i=1/2$ is genuinely entangled. As previously, if the vector $(\lambda_0,\ldots,\lambda_M)$ is outside the sphere $S$ centered at $C=(c_0,\ldots,c_M)$ with $c_i=\binom{N}{i}2^{-(2i+1)}$ and going through the origin, then the state \eqref{decomprho} with $\ket{\psi_i}$ given by \eqref{psiDicke} is genuinely entangled. The intersection of the outside of the sphere with the region $\lambda_i\geq 0$ and the plane $\sum_{i=0}^{M}\lambda_i=1/2$ is nonempty. Indeed, the point $E=(0,\ldots,0,1/2)\in \bbr^M$ lies strictly outside the sphere if and only if $c_{M}<1/4$, which is the case since $\binom{2M+1}{M}<4^M$ for all $M\geq1$. The distance $d$ from $E$ to the sphere constrained by the condition $\sum_{i=0}^{M}\lambda_i=1/2$ can be obtained by introducing Lagrange multipliers. All $\lambda_i$ with $\lambda_i\geq 0$, $\sum_{i=0}^{M}\lambda_i=1/2$ and at a distance less than $d$ from $E$ give genuinely entangled states.

For $N=3$, the point on the sphere $S$ and the plane $\lambda_0+\lambda_1=1/2$ which is closest to $E=(0,1/2)$ is $(\lambda_0,\lambda_1)=(1/16,7/16)$. Thus all states with $\lambda_1>7/16$ are genuinely entangled. In fact, in this case, we have the necessary and sufficient condition of Sec.~\ref{subsec:3qubit} which reads $A\geq 0$. Using Eq.~(\ref{xab}), this is equivalent to $\lambda_0\geq \lambda_1/3\geq 0$. Thus the state is genuinely entangled if and only if $\lambda_1>3/8$. In Fig.~\ref{fig2} (a), the region outside the red solid curve corresponds to genuinely entangled states detected by the criterion (\ref{genuine}), while states outside the blue dashed circle correspond to those detected by the analytical bound (\ref{eqn:sphere}). The red solid curve intersects the line $\lambda_0+\lambda_1=1/2$ at $\lambda_1=3/8$, which coincides with our necessary and sufficient condition.

For higher $N$, the bound (\ref{eqn:sphere}) allows us to obtain closed analytical expressions for a region of admissible values of $\lambda_i$. A visualization of the case $N=5$ can be found in Fig.~\ref{fig2} (b), where states outside the red curve are obtained numerically from the criterion (\ref{genuine}), and states outside the blue dashed circle correspond to the analytical bound (\ref{eqn:sphere}).

%%%%%%%%%%%%%%%%%%%%%%%%%%%%%%%%%%%%%%%%%%%%%%%%%%%%%%%%%%%%%%%%%%%%%%%%%
%%%%%%%%%%%%%%%%%%%%%%%%%%%%%%%%%%%%%%%%%%%%%%%%%%%%%%%%%%%%%%%%%%%%%%%%%
\section{Conclusions}
%%%%%%%%%%%%%%%%%%%%%%%%%%%%%%%%%%%%%%%%%%%%%%%%%%%%%%%%%%%%%%%%%%%%%%%%%
%%%%%%%%%%%%%%%%%%%%%%%%%%%%%%%%%%%%%%%%%%%%%%%%%%%%%%%%%%%%%%%%%%%%%%%%%

In this paper, we have investigated genuinely entangled states in the set $\nocorset_N$ of $N$-qubit symmetric states with no $N$-partite correlations. The tensor representation~\cite{Gir15} has allowed us to give a simple characterization of these states. From this characterization, it easily follows that no such states exist for an even number of qubits. We have shown that for any $\rho\in\nocorset_N$ with $N$ odd, all eigenspaces of $\rho$ are even-dimensional, and that the general form of the state is given by (\ref{decomprho}). This form generalizes the mixture of a state with its antistate investigated in~\cite{Sch15}. The parametrization (\ref{rho3}) has allowed us to find a simple necessary separability condition for $\rho\in\nocorset_N$, namely $A\geq 0$ with $A=(\langle \sigma_{ab}\rangle_\rho)_{1\leq a,b\leq 3}$ and $\sigma_{ab}=\sigma_a\otimes\sigma_b\otimes \mathbb{1}_2\otimes \cdots\otimes \mathbb{1}_2$.

In the case of three qubits, we have shown that $A\geq 0$ is in fact a necessary and sufficient condition for a state in $\nocorset_3$ to be separable. Interestingly, this leads to the equivalence between separability of $\rho\in\nocorset_3$ and separability of its two-qubit reduced density matrix. This implies that the entanglement in states $\rho\in\nocorset_3$ cannot be entirely destroyed upon the loss of one qubit. 

In the case of rank-2 states $\rho\in\nocorset_N$, we have obtained a necessary and sufficient condition for separability in terms of the same matrix $A$ as $\mathrm{rank}A=1$. This condition has allowed us to generalize the construction to families of arbitrary rank, by considering mixtures in the vicinity of rank-2 genuinely entangled $\rho\in\nocorset_N$.

\section*{Acknowledgments}
O.G. thanks the University of Li\`ege for hospitality.


\begin{thebibliography}{99}
\bibitem{Wer89} F.~Werner, {\it Quantum states with Einstein-Podolsky-Rosen correlations admitting a hidden-variable model}, Phys.~Rev.~A {\bf 40}, 4277 (1989).

\bibitem{Ben11} C.~H.~Bennett, A.~Grudka, M.~Horodecki, P.~Horodecki, and R.~Horodecki, {\it Postulates for measures of genuine multipartite correlations}, Phys. Rev. A {\bf 83}, 012312 (2011).

\bibitem{Oll01} H.\ Ollivier and W.\ H.\ Zurek, {\it Quantum Discord: A Measure of the Quantumness of Correlations}, Phys.\ Rev.\ Lett.\ \textbf{88}, 017901 (2001).

\bibitem{Hen01} L.\ Henderson and V.\ Vedral, {\it Classical, quantum and total correlations}, J.\ Phys.\ A: Math.\ Gen.\ \textbf{34}, 6899 (2001). 

\bibitem{Kas08} D.~Kaszlikowski, A.\ S.\ De, U.\ Sen, V.\ Vedral, and A. Winter, {\it Quantum Correlation without Classical Correlations}, Phys.~Rev.~Lett.~{\bf 101}, 070502 (2008).

\bibitem{Wie09} W.\ Wieczorek, R.\ Krischek, N.\ Kiesel, P.\ Michelberger, G.\ T\'oth, and H.\ Weinfurter, {\it Experimental Entanglement of a Six-Photon Symmetric Dicke State}, Phys.~Rev.~Lett.~{\bf 103}, 020504 (2009).

\bibitem{Sch15} C. Schwemmer, L. Knips, M. C. Tran, A. de Rosier, W. Laskowski, T. Paterek, and H. Weinfurter, {\it Genuine multipartite entanglement without multipartite
correlations}, Phys.~Rev.~Lett.~{\bf 114}, 180501 (2015).

\bibitem{Wie12} M.~Wie\'sniak, M.~Nawareg, and M.~\.Zukowski, {\it N-particle nonclassicality without N-particle correlations}, Phys. Rev. A {\bf 86}, 042339 (2012).

\bibitem{Tra17} M.~C.~Tran, M.~Zuppardo, A.~de~Rosier, L.~Knips, W.~Laskowski, T.~Paterek, and H.~Weinfurter, {\it Genuine $N$-partite entanglement without $N$-partite correlation functions}, Phys.~Rev.~A~{\bf 95}, 062331 (2017).

\bibitem{Gir15} O.~Giraud, D.~Braun, D.~Baguette, T.~Bastin, and J.~Martin, {\it Tensor representation of spin states}, Phys.~Rev.~Lett.~{\bf 114}, 080401 (2015).

\bibitem{Boh16} F.~Bohnet-Waldraff, D.~Braun, and O.~Giraud, {\it Partial transpose criteria for symmetric states}, Phys.~Rev.~A \textbf{94}, 042343 (2016).

\bibitem{Ich08} T.~Ichikawa, T.~Sasaki, I.~Tsutsui, and N.~Yonezawa, {\it Exchange symmetry and multipartite entanglement}, Phys.~Rev.~A \textbf{78}, 052105 (2008).

\bibitem{Las12} W.\ Laskowski, M.\ Markiewicz, T.\ Paterek and M.\ Wie\'{s}niak, {\it Incompatible local hidden-variable models of quantum correlations}, Phys.~Rev.~A~{\bf 86}, 032105 (2012).

\bibitem{Bag14}
D.~Baguette, J.~Martin, and T.~Bastin {\it Multiqubit symmetric states with maximally mixed one-qubit reductions}, Phys.~Rev.~A~{\bf 90}, 032314 (2014).

\bibitem{Buz99} V.~Bu\v{z}ek, M.~Hillery, and R.~F.~Werner, {\it Optimal manipulations with qubits: Universal-NOT gate}, Phys.~Rev.~A {\bf 60}, R2626(R) (1999).

\bibitem{Bad08} P.~Badzi\c{a}g, \v{C}.~Brukner, W.~Laskowski, T.~Paterek, and M.~\.{Z}ukowski, {\it Experimentally Friendly Geometrical Criteria for Entanglement}, Phys.~Rev.~Lett.~{\bf 100}, 140403 (2008).

\bibitem{Hor96} M.~Horodecki, P.~Horodecki, and R.~Horodecki, {\it Separability of mixed states: necessary and sufficient conditions}, Phys.~Lett.~A \textbf{223}, 1 (1996).

\bibitem{Aug12} R.~Augusiak, J.~Tura, J.~Samsonowicz, and M.~Lewenstein, {\it Entangled symmetric states og $N$ qubits with all positive partial transpositions}, Phys. Rev. A {\bf 86}, 042316 (2012).

\bibitem{Mar13} M.~Markiewicz, W.~Laskowski, T.~Paterek, and M.~\.Zukowski, {\it Detecting genuine multipartite entanglement of pure states with bipartite correlations}, Phys. Rev. A {\bf 87}, 034301 (2013).


\end{thebibliography}
\end{document}